\documentclass[aoas,nameyear,seceqn,mtplusscr,dvips]{arximspdf}
\usepackage{mathrsfs,mathbh,multirow}
\usepackage{rotating}
\usepackage{graphicx}
\doi{10.1214/09-AOAS309}
\volume{4}
\issue{1}
\pubyear{2010}
\firstpage{124}
\lastpage{150}


\begin{document}
\begin{frontmatter}

\title{Hierarchical relational models for~document~networks}
\runtitle{Hierarchical relational models for document networks}

\begin{aug}
\author[A]{\fnms{Jonathan} \snm{Chang}\thanksref{a1}\ead[label=e1]{jonchang@facebook.com}\corref{}} \and
\author[B]{\fnms{David M.} \snm{Blei}\thanksref{a2}\ead[label=e2]{blei@cs.princeton.edu}}
\runauthor{J. Chang and D. M. Blei}
\thankstext{a1}{Work done while at Princeton University.}
\thankstext{a2}{Supported by ONR 175-6343, NSF CAREER
0745520 and grants from Google and Microsoft.}
\affiliation{Facebook and Princeton University}
\address[A]{Facebook \\
1601 S California Ave. \\
Palo Alto, California 94304 \\USA\\
\printead{e1}} 
\address[B]{Department of Computer Science\\
Princeton University\\
34 Olden St.\\
Princeton, New Jersey 08544 \\USA\\
\printead{e2}}
\end{aug}

\received{\smonth{1} \syear{2009}}
\revised{\smonth{11} \syear{2009}}

%
\begin{abstract}
We develop the relational topic model (RTM), a hierarchical model of
both network structure and node attributes. We focus on document
networks, where the attributes of each document are its words, that is,
discrete observations taken from a fixed vocabulary. For each pair
of documents, the RTM models their link as a binary random variable
that is conditioned on their contents. The~model can be used to
summarize a network of documents, predict links between them, and
predict words within them. We derive efficient inference and
estimation algorithms based on variational methods that take
advantage of sparsity and scale with the number of links. We
evaluate the predictive performance of the RTM for large networks of
scientific abstracts, web documents, and geographically tagged news.
\end{abstract}

%
\begin{keyword}
\kwd{Mixed-membership models}
\kwd{variational methods}
\kwd{text analysis}
\kwd{network models}.
\end{keyword}

\end{frontmatter}
%

\section{Introduction}

Network data, such as citation networks of documents, hyperlinked
networks of web pages, and social networks of friends, are pervasive
in applied statistics and machine learning. The~statistical analysis
of network data can provide both useful predictive models and
descriptive statistics. Predictive models can point social network
members toward new friends, scientific papers toward relevant
citations, and web pages toward other related pages. Descriptive
statistics can uncover the hidden community structure underlying a
network data set.

Recent research in this field has focused on latent variable models of
link structure, models that decompose a network according to hidden
patterns of connections between its nodes [\citet{Kemp2004p2068}; \citet{Hofman2007p1340}; \citet{Airoldi2007p1863}]. These models
represent a significant departure from statistical models of networks,
which explain network data in terms of observed sufficient
statistics [\citet{Fienberg1985p8582}; \citet{Wasserman1996p823}; \citet{Getoor2001p983}; \citet{Newman2002p7627}; \citet{Taskar2004p820}].

While powerful, current latent variable models account only for the
structure of the network, ignoring additional attributes of the nodes
that might be available. For example, a citation network of articles
also contains text and abstracts of the documents, a linked set of
web-pages also contains the text for those pages, and an on-line
social network also contains profile descriptions and other
information about its members. This type of information about the
nodes, along with the links between them, should be used for
uncovering, understanding, and exploiting the latent structure in the
data.

To this end, we develop a new model of network data that accounts for
both links and attributes. While a traditional network model requires
some observed links to provide a predictive distribution of links for
a node, our model can predict links using only a new node's
attributes. Thus, we can suggest citations of newly written papers,
predict the likely hyperlinks of a web page in development, or suggest
friendships in a social network based only on a new user's profile of
interests. Moreover, given a new node and its links, our model
provides a predictive distribution of node attributes. This mechanism
can be used to predict keywords from citations or a user's interests
from his or her social connections. Such prediction problems are out
of reach for traditional network models.

Here we focus on document networks. The~attributes of each document are
its text, that is, discrete observations taken from a fixed vocabulary,
and the links between documents are connections such as friendships,
hyperlinks, citations, or adjacency. To model the text, we build on
previous research in mixed-membership document models, where each
document exhibits a latent mixture of multinomial distributions or
``topics'' [\citet{Blei2003p1135}; \citet{Erosheva2004p8486}; \citet{Steyvers2007p6822}].
The~links are then modeled dependent on this latent representation.
We call our model, which explicitly ties the content of the documents
with the connections between them, the \textit{relational topic
model} (RTM).

The~RTM affords a significant improvement over previously developed
models of document networks. Because the RTM jointly models node
attributes and link structure, it can be used to make predictions
about one given the other. Previous work tends to explore one or the
other of these two prediction problems. Some previous work uses link
structure to make attribute predictions [\citet{Chakrabarti1998p1575}; \citet{Kleinberg1999p1367}], including several topic
models [\citet{McCallum2005p1338}; \citet{Wang2005p1342}; \citet{Dietz2007p181}].
However, none of these methods can make predictions about links given
words.

Other models use node attributes to predict
links [\citet{Hoff2002p1787}]. However, these models condition on the
attributes but do not model them. While this may be effective for
small numbers of attributes of low dimension, these models cannot make
meaningful predictions about or using high-dimensional attributes such
as text data. As our empirical study in Section \ref{sec:results}
illustrates, the mixed-membership component provides dimensionality
reduction that is essential for effective prediction.

In addition to being able to make predictions about links given words
and words given links, the RTM is able to do so for \textit{new}
documents---documents outside of the training data. Approaches which
generate document links through topic models treat links as discrete
``terms'' from a separate vocabulary that essentially indexes the
observed documents [\citet{Cohn2001p1231}; \citet{Erosheva2004p8486}; \citet{Gruber2008p6856}; \citet{Nallapati2008p1331};
\citet{Sinkkonen2008p5772}].
Through this index, such approaches encode the observed training data
into the model and thus cannot generalize to observations outside of
them. Link and word predictions for new documents, of the kind we
evaluate in Section~\ref{sec:eval-prediction}, are ill defined.

\citeauthor{Xu2006} (\citeyear{Xu2006}, \citeyear{Xu2008}) have jointly modeled links and document content
using nonparametric Bayesian techniques so as to avoid these
problems. However, their work does not assume mixed-memberships, which
have been shown to be useful for both document
modeling [\citet{Blei2003p1135}] and network
modeling [\citet{Airoldi2007p1863}]. Recent work from
\citet{Nallapati2008p7255} has also jointly modeled links and document
content. We elucidate the subtle but important differences between
their model and the RTM in Section~\ref{sec:lpf}. We then demonstrate in
Section~\ref{sec:eval-prediction} that the RTM makes modeling assumptions
that lead to significantly better predictive performance.

The~remainder of this paper is organized as follows. First, we
describe the statistical assumptions behind the relational topic
model. Then, we derive efficient algorithms based on variational
methods for approximate posterior inference, parameter estimation, and
prediction. Finally, we study the performance of the RTM on
scientific citation networks, hyperlinked web pages, and
geographically tagged news articles. The~RTM provides better word
prediction and link prediction than natural alternatives and the
current state of the art.

\section{Relational topic models}
\label{sec:model}

\begin{figure}

\includegraphics{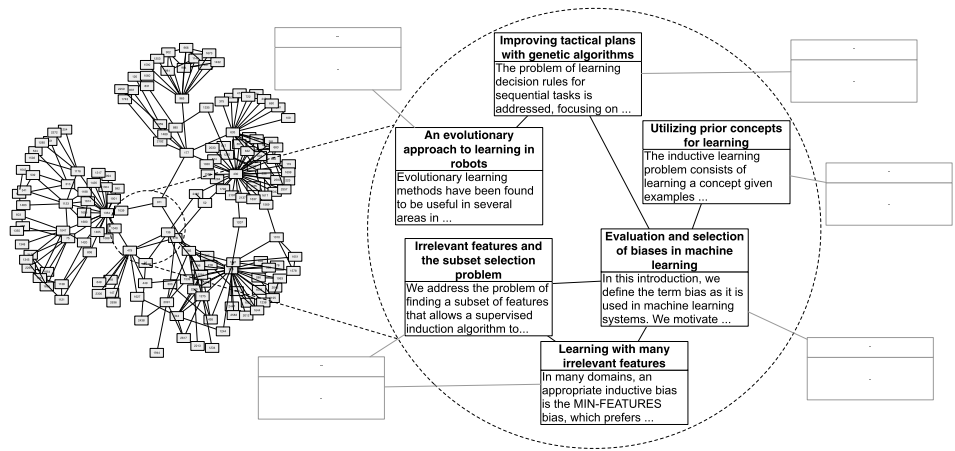}

\caption{Example data appropriate for the relational topic model.
Each document is represented as a bag of words and linked to other
documents via citation. The~RTM defines a joint distribution over
the words in each document and the citation links between them.}
\label{figure:data-example}
\end{figure}

The~\textit{relational topic model} (RTM) is a hierarchical
probabilistic model of networks, where each node is endowed with
attribute information. We will focus on text data, where the
attributes are the words of the documents (see
Figure~\ref{figure:data-example}). The~RTM embeds this data in a latent
space that explains both the words of the documents and how they are
connected.

\subsection{Modeling assumptions}

The~RTM builds on previous work in
mixed-membership document models. Mixed-membership models are latent
variable models of heterogeneous data, where each data point can
exhibit multiple latent components. Mixed-membership models have been
successfully applied in many domains, including survey
data [\citet{Erosheva2007}], image
data [\citet{Barnard2003}; \citet{FeiFei2005p8314}], rank
data [\citet{Gormley2009lq}], network data [\citet
{Airoldi2007p1863}] and
document modeling [\citet{Blei2003p1135}; \citet{Steyvers2007p6822}].
Mixed-membership models were independently developed in the field of
population genetics [\citet{Pritchard2000}].


\setcounter{footnote}{2}
To model node attributes, the RTM reuses the statistical assumptions
behind latent Dirichlet allocation (LDA) [\citet{Blei2003p1135}], a
mixed-membership model of documents.\footnote{A general
mixed-membership model can accommodate any kind of grouped data
paired with an appropriate observation
model [\citet{Erosheva2004p8486}].} Specifically, LDA is a
hierarchical probabilistic model that uses a set of ``topics,''
distributions over a fixed vocabulary, to describe a corpus of
documents. In its generative process, each document is endowed with
a Dirichlet-distributed vector of topic proportions, and each word
of the document is assumed drawn by first drawing a topic assignment
from those proportions and then drawing the word from the
corresponding topic distribution. While a traditional mixture model
of documents assumes that every word of a document arises from a
single mixture component, LDA allows each document to exhibit
multiple components via the latent topic proportions vector.

In the RTM, each document is first generated from topics as in LDA.
The~links between documents are then modeled as binary variables, one
for each pair of documents. These binary variables are distributed
according to a distribution that depends on the topics used to
generate each of the constituent documents. Because of this
dependence, the content of the documents is statistically connected
to the link structure between them. Thus, each document's
mixed-membership depends both on the content of the document as well
as the pattern of its links. In turn, documents whose memberships are
similar will be more likely to be connected under the model.

\begin{figure}[b]

\includegraphics{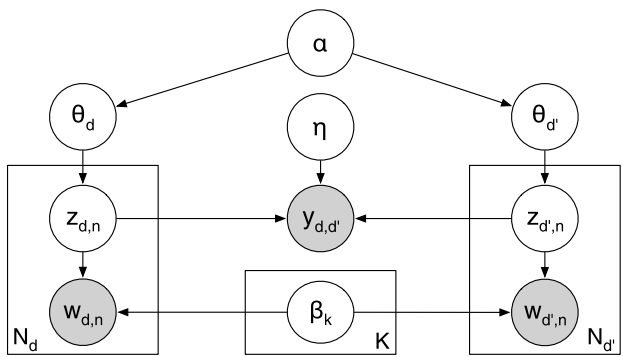}

\caption{A two-document segment of the RTM. The~variable
$y_{d,d'}$ indicates whether the two documents are linked.
The~complete model contains this variable for each pair of documents.
This binary variable is generated contingent on the topic
assignments for the participating documents, $\mathbf{z}_d$~and~$\mathbf{z}_{d'}$,
and global regression parameters $\bolds{\eta}$. The~plates
indicate replication. This model captures both the words and the
link structure of the data shown in Figure~\protect\ref{figure:data-example}.}
\label{figure:slda-link}
\end{figure}

The~parameters of the RTM are as follows: the topics $\bolds{\beta
}_{1:K}$, $K$
multinomial parameters each describing a distribution on words; a
$K$-dimensional Dirichlet parameter $\alpha$; and a function $\psi$
that provides binary probabilities. (This function is explained in
detail below.) We denote a set of observed documents by $w_{1:D,
1:N}$, where $w_{i, 1:N}$ are the words of the $i$th document.
(Words are assumed to be discrete observations from a fixed
vocabulary.) We denote the links between the documents as binary
variables $y_{1:D, 1:D}$, where $y_{i,j}$ is one if there is a link
between the $i$th and $j$th document. The~RTM assumes that a set of
observed documents $w_{1:D, 1:N}$ and binary links between them
$y_{1:D, 1:D}$ are generated by the following process:
\begin{enumerate}
\item For each document $d$:
\begin{enumerate}
\item[(a)] Draw topic proportions $\theta_d|\alpha\sim
\operatorname{Dir}(\alpha).$
\item[(b)] For each word $w_{d,n}$:
\begin{enumerate}
\item[i.] Draw assignment $z_{d,n}|\theta_d \sim\operatorname{Mult}
(\theta_d)$.
\item[ii.] Draw word $w_{d,n}|z_{d,n}, \bolds{\beta}_{1:K} \sim
\operatorname{Mult}(\bolds{\beta}_{z_{d,n}}).$
\end{enumerate}
\end{enumerate}
\item For each pair of documents $d$, $d'$:
\begin{enumerate}
\item[(a)] Draw binary link indicator
\[
y_{d,d'}|\mathbf{z}_d, \mathbf{z}_{d'} \sim
\psi(\cdot| \mathbf{z}_{d}, \mathbf{z}_{d'}, \bolds{\eta}),
\]
where $\mathbf{z}_d = \{ z_{d,1}, z_{d,2}, \ldots, z_{d,n} \}$.
\end{enumerate}
\end{enumerate}
Figure~\ref{figure:slda-link} illustrates the graphical model for this
process for a single pair of documents. The~full model, which is
difficult to illustrate in a small graphical model, contains the
observed words from all $D$ documents, and $D^2$ link variables for
each possible connection between them.

%

\subsection{Link probability function}
\label{sec:lpf}
The~function $\psi$ is the \textit{link probability function} that
defines a distribution over the link between two documents. This
function is dependent on the two vectors of topic assignments that
generated their words, $\mathbf{z}_d$ and $\mathbf{z}_{d'}$.

This modeling decision is important. A natural alternative is to
model links as a function of the topic proportions vectors
$\theta_d$ and $\theta_{d'}$. One such model is that
of \citet{Nallapati2008p7255}, which extends the mixed-membership
stochastic blockmodel [\citet{Airoldi2007p1863}] to generate node
attributes. Similar in spirit is the nongenerative model of
\citet{Mei2008p6870} which ``regularizes'' topic models with graph
information. The~issue with these formulations is that the links and
words of a single document are possibly explained by disparate sets of
topics, thereby hindering their ability to make predictions about
words from links and vice versa.

In enforcing that the link probability function depends on the latent
topic assignments $\mathbf{z}_d$ and $\mathbf{z}_{d'}$, we enforce that the
specific topics used to generate the links are those used to generate
the words. A similar mechanism is employed in \citet{Blei2007p105}
for nonpair-wise response variables. In estimating parameters, this
means that the same topic indices describe both patterns of recurring
words and patterns in the links. The~results in
Section~\ref{sec:eval-prediction} show that this provides a superior
prediction mechanism.

We explore four specific possibilities for the link probability
function. First, we consider
%
\begin{equation}
\label{eq:psisigma}
\psi_{\sigma}(y = 1) = \sigma\bigl({\bolds{\eta}}^{\mathrm{T}}
(\overline{\mathbf{z}}_d \circ\overline{\mathbf{z}}_{d'}) + \nu\bigr),
\end{equation}
where $\overline{\mathbf{z}}_{d} = \frac{1}{N_d}\sum_n z_{d,n}$,
the $\circ$
notation denotes the Hadamard (element-wise)\vspace*{1pt} product, and the function
$\sigma$ is the sigmoid. This link function models each per-pair
binary variable as a logistic regression with hidden covariates. It
is parameterized by coefficients $\eta$ and intercept $\nu$. The~covariates are constructed by the Hadamard product of $\overline
{\mathbf{z}}_d$
and $\overline{\mathbf{z}}_{d'}$, which captures similarity between
the hidden
topic representations of the two documents.

Second, we consider
%
\begin{equation}
\label{eq:psie}
\psi_{e}(y = 1) = \exp\bigl({\bolds{\eta}}^{\mathrm{T}} (\overline
{\mathbf{z}}_d \circ
\overline{\mathbf{z}}_{d'}) + \nu\bigr).
\end{equation}
Here, $\psi_{e}$ uses the same covariates as $\psi_{\sigma}$, but has
an exponential mean function instead. Rather than tapering off when
$\overline{\mathbf{z}}_d$ and $\overline{\mathbf{z}}_{d'}$ are
close (i.e., when their
weighted inner product, ${\bolds{\eta}}^{\mathrm{T}} (\overline
{\mathbf{z}}_d \circ
\overline{\mathbf{z}}_{d'})$, is large), the probabilities returned
by this
function continue to increase exponentially. With some algebraic
manipulation, the function $\psi_{e}$ can be viewed as an approximate
variant of the modeling methodology presented in
\citet{Blei2003p1754}.

Third, we consider
%
\begin{equation}
\label{eq:psiphi}
\psi_{\Phi}(y = 1) = \Phi\bigl({\bolds{\eta}}^{\mathrm{T}} (\overline
{\mathbf{z}}_d
\circ
\overline{\mathbf{z}}_{d'}) + \nu\bigr),
\end{equation}
where $\Phi$ represents the cumulative distribution function of the
Normal distribution. Like $\psi_\sigma$, this link function models
the link response as a regression parameterized by coefficients $\eta$
and intercept $\nu$. The~covariates are also constructed by the
Hadamard product of $\overline{\mathbf{z}}_d$ and $\overline{\mathbf
{z}}_{d'}$, but instead
of the logit model hypothesized by $\psi_\sigma$, $\psi_\Phi$ models
the link probability with a probit model.

Finally, we consider
%
\begin{equation}
\label{eq:psiN}
\psi_{N}(y = 1) = \exp\bigl(-{\bolds{\eta}}^{\mathrm{T}} (\overline
{\mathbf{z}}_d
- \overline{\mathbf{z}}_{d'}) \circ(\overline{\mathbf{z}}_d -
\overline{\mathbf{z}}_{d'}) - \nu
\bigr).
\end{equation}
Note that $\psi_N$ is the only one of the link probability functions
which is not a function of $\overline{\mathbf{z}}_d \circ\overline
{\mathbf{z}}_{d'}$.
Instead, it depends on a weighted squared Euclidean difference between
the two latent topic assignment distributions. Specifically, it is
the multivariate Gaussian density function, with mean 0 and diagonal
covariance characterized by $\bolds{\eta}$, applied to $\overline
{\mathbf{z}}_d -
\overline{\mathbf{z}}_{d'}$. Because the range of $\overline{\mathbf
{z}}_d -
\overline{\mathbf{z}}_{d'}$ is finite, the probability of a link,
$\psi_{N}(y=1)$, is also finite. We constrain the parameters
$\bolds{\eta}$ and $\nu$ to ensure that it is between zero and one.

\begin{figure}

\includegraphics{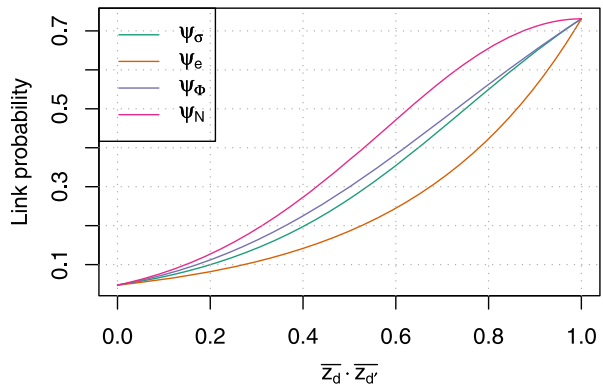}

\caption{A comparison of different link probability functions.
The~plot shows the probability of two documents being linked as a
function of their similarity (as measured by the inner product of
the two documents' latent topic assignments). All link
probability functions were parameterized so as to have the same
endpoints.}
\label{figure:lpfs}
\end{figure}

All four of the $\psi$ functions we consider are plotted in
Figure~\ref{figure:lpfs}. The~link likelihoods suggested by the link
probability functions are plotted against the inner product of
$\overline{\mathbf{z}}_d$ and $\overline{\mathbf{z}}_{d'}$. The~parameters of the link
probability functions were chosen to ensure that all curves have the
same endpoints. Both $\psi_\sigma$ and $\psi_\Phi$ have similar
sigmoidal shapes. In contrast, the $\psi_e$ is exponential in shape
and its slope remains large at the right limit. The~one-sided
Gaussian form of $\psi_N$ is also apparent.

\section{Inference, estimation and prediction}

With the model defined, we turn to approximate posterior inference,
parameter estimation, and prediction. We develop a variational
inference procedure for approximating the posterior. We use this
procedure in a variational expectation-maximization (EM) algorithm for
parameter estimation. Finally, we show how a model whose parameters
have been estimated can be used as a predictive model of words and
links.

\subsection{Inference}
\label{sec:inference}

The~goal of posterior inference is to compute the posterior
distribution of the latent variables conditioned on the observations.
As with many \mbox{hierarchical} Bayesian models of interest, exact posterior
inference is intractable and we appeal to approximate inference
methods. Most previous work on latent variable network modeling has
employed Markov Chain Monte Carlo (MCMC) sampling methods to
approximate the posterior of
interest [\citet{Hoff2002p1787}; \citet{Kemp2004p2068}]. Here, we employ
variational inference [\citet{Jordan1999p13}; \citet{Wainwright2005}], a
deterministic alternative to MCMC sampling that has been shown to give
comparative accuracy to MCMC with improved computational
efficiency [\citet{Blei2006p116}; \citet{Braun2007p7339}].
\citet{Wainwright2009} investigate the properties of variational
approximations in detail. Recently, variational methods have been
employed in other latent variable network
models [\citet{Hofman2007p1340}; \citet{Airoldi2007p1863}].

In variational methods, we posit a family of distributions over the
latent variables, indexed by free variational parameters. Those
parameters are then fit to be close to the true posterior, where
closeness is measured by relative entropy. For the RTM, we use the
fully-factorized family, where the topic proportions and all topic
assignments are considered independent,
%
\begin{equation}
q(\bolds{\Theta}, \mathbf{Z} |\bolds{\gamma}, \bolds{\Phi}) =
\prod_d \biggl[ q_\theta(\theta_d | \gamma_d) \prod_n q_z(z_{d,n} |
\phi_{d,n})\biggr].
\label{eqn:mean-field}
\end{equation}
The~parameters $\bolds{\gamma}$ are variational Dirichlet parameters, one
for each document, and~$\bolds{\Phi}$ are variational multinomial
parameters, one for each word in each document. Note that
$\mathbb{E}_{q}[z_{d,n}] = \phi_{d,n}$.


Minimizing the relative entropy is equivalent to maximizing the
Jensen's lower bound on the marginal probability of the observations,
that is, the evidence lower bound (ELBO),
\begin{eqnarray}\label{eqn:posterior}
\hspace*{18pt}\mathscr{L} &=& \sum_{(d_1, d_2)} \mathbb{E}_{q}[\log p(y_{d_1,d_2}
| \mathbf{z}_{d_1}, \mathbf{z}_{d_2}, \bolds{\eta}, \nu)] + \sum_d \sum
_n \mathbb{E}_{q}[\log p(z_{d,n} | \bolds{\theta}_d)]\hspace*{-18pt} \nonumber\\[-8pt]\\[-8pt]
\hspace*{18pt}& &{} +\sum_d \sum_n \mathbb{E}_{q}[\log p(w_{d, n} | \bolds{\beta
}_{1:K}, z_{d, n})] +
\sum_d \mathbb{E}_{q}[\log p(\bolds{\theta}_d | \alpha)] +
\mathrm{H}(q),\hspace*{-18pt}\nonumber
\end{eqnarray}
where $(d_1, d_2)$ denotes all document pairs and $\mathrm{H}(q)$
denotes the entropy of the distribution $q$. The~first term of the
ELBO differentiates the RTM from LDA [\citet{Blei2003p1135}]. The~connections between documents affect the objective in approximate
posterior inference (and, below, in parameter estimation).

We develop the inference procedure below under the assumption that
only observed links will be modeled (i.e., $y_{d_1, d_2}$ is either
$1$ or unobserved).\footnote{Sums over document pairs $(d_1, d_2)$ are
understood to range over pairs for which a link has been observed.}
We do this for both methodological and computational reasons.

First, while one can fix $y_{d_1, d_2} = 1$ whenever a link is
observed between $d_1$ and~$d_2$ and set $y_{d_1, d_2} = 0$ otherwise,
this approach is inappropriate in corpora where the absence of a link
cannot be construed as evidence for $y_{d_1, d_2} = 0$. In these
cases, treating these links as unobserved variables is more faithful
to the underlying semantics of the data. For example, in large social
networks such as Facebook the absence of a link between two people
does not necessarily mean that they are not friends; they may be real
friends who are unaware of each other's existence in the network.
Treating this link as unobserved better respects our lack of knowledge
about the status of their relationship.


Second, treating nonlinks links as hidden decreases the computational
cost of inference; since the link variables are leaves in the
graphical model, they can be removed whenever they are unobserved.
Thus, the complexity of computation scales linearly with the number of
observed links rather than the number of document pairs. When the
number of true observations is sparse relative to the number of
document pairs, as is typical, this provides a significant
computational advantage. For example, on the \textit{Cora} data set
described in Section~\ref{sec:results}, there are 3,665,278 unique document
pairs but only 5278 observed links. Treating nonlinks as hidden in
this case leads to an inference procedure which is nearly 700 times
faster.

Our aim now is to compute each term of the objective function given in
equation~(\ref{eqn:posterior}). The~first term,
%
\begin{equation}
\sum_{(d_1, d_2)} \mathscr{L}_{d_1, d_2} \equiv\sum_{(d_1, d_2)}
\mathbb{E}_{q}[\log p(y_{d_1,d_2} | \mathbf{z}_{d_1}, \mathbf{z}_{d_2},
\bolds{\eta}, \nu)],
\label{eqn:firstterm}
\end{equation}
depends on our choice of link probability function. For many link
probability functions, this term cannot be expanded analytically.
However, if the link probability function depends only on
$\overline{\mathbf{z}}_{d_1} \circ\overline{\mathbf{z}}_{d_2}$, we
can expand the
expectation using the following first-order
approximation [\citet{Braun2007p7339}]\footnote{While we do not give a
detailed proof here, the error of a first-order approximation is
closely related to the probability mass in the tails of the
distribution on $\overline{\mathbf{z}}_{d_1}$ and $\overline{\mathbf
{z}}_{d_2}$. Because
the number of words in a document is typically large, the variance of
$\overline{\mathbf{z}}_{d_1}$ and $\overline{\mathbf{z}}_{d_2}$
tends to be small, making
the first-order approximation a good one.}:
\[
\mathscr{L}_{(d_1, d_2)} =
\mathbb{E}_{q}[\log\psi(\overline{\mathbf{z}}_{d_1} \circ
\overline{\mathbf{z}}_{d_2})] \approx
\log\psi(\mathbb{E}_{q}[\overline{\mathbf{z}}_{d_1} \circ
\overline{\mathbf{z}}_{d_2}]) =
\log\psi(\overline{\bolds{\pi}}_{d_1, d_2}),
\]
where $\overline{\bolds{\pi}}_{d_1, d_2}= \overline{\bolds{\phi
}}_{d_1} \circ\overline{\bolds{\phi}}_{d_2}$ and
$\overline{\bolds{\phi}}_{d} = \mathbb{E}_{q}[\overline{\mathbf
{z}}_{d}] = \frac{1}{N_d}\sum_n
\phi_{d,n}$.
In this work, we explore three functions which can be written in this
form,
%
\begin{eqnarray}\label{eq:approxexpect}
\mathbb{E}_{q}[\log\psi_\sigma(\overline{\mathbf{z}}_{d_1} \circ
\overline{\mathbf{z}}_{d_2})]
&\approx&\log\sigma({\bolds{\eta}}^{\mathrm{T}} \overline{\bolds
{\pi}}_{d_1, d_2}+ \nu),
\nonumber\\
\mathbb{E}_{q}[\log\psi_\Phi(\overline{\mathbf{z}}_{d_1} \circ
\overline{\mathbf{z}}_{d_2})]
&\approx&\log\Phi({\bolds{\eta}}^{\mathrm{T}} \overline{\bolds{\pi
}}_{d_1, d_2}+ \nu), \\
\mathbb{E}_{q}[\log\psi_e(\overline{\mathbf{z}}_{d_1} \circ
\overline{\mathbf{z}}_{d_2})] &=&
{\bolds{\eta}}^{\mathrm{T}} \overline{\bolds{\pi}}_{d_1, d_2}+ \nu.
\nonumber
\end{eqnarray}
Note that for $\psi_e$ the expression is exact. The~likelihood when
$\psi_N$ is chosen as the link probability function can also be
computed exactly,
\[
\mathbb{E}
_{q}[\log\psi_N(\overline{\mathbf{z}}_{d_1}, \overline
{\mathbf{z}}_{d_2})] = -\nu-
\sum_i \eta_i\bigl(
(\overline{\bolds{\phi}}_{d_1,i} - \overline{\bolds{\phi
}}_{d_2,i})^2 + \operatorname{Var}(\overline{\mathbf{z}}
_{d_1,i}) + \operatorname{Var}(\overline{\mathbf{z}}_{d_2,i})\bigr),
\]
where $\overline{\mathbf{z}}_{d,i}$ denotes the $i$th element of the
mean topic
assignment vector, $\overline{\mathbf{z}}_d$, and $\operatorname
{Var}(\overline{\mathbf{z}}_{d,i}) =
\frac{1}{N_d^2} \sum_n \phi_{d,n,i} (1 - \phi_{d,n,i})$, where
$\phi_{d,n,i}$ is the $i$th element of the multinomial parameter
$\phi_{d,n}$. (See Appendix~\ref{app:ascent}.)

Leveraging these expanded expectations, we then use coordinate ascent
to optimize the ELBO with respect to the variational parameters
$\bolds{\gamma}, \bolds{\Phi}$. This yields an approximation to the true
posterior. The~update for the variational multinomial $\phi_{d,j}$ is
%
\begin{equation}
\phi_{d,j} \propto \exp\biggl\{\sum_{d' \ne d} \nabla_{\phi
_{d,n}} \mathscr{L}_{d, d'}+
\mathbb{E}_{q}[\log\theta_d | \gamma_d] + \log\bolds{\beta}_{\cdot,
w_{d,j}}\biggr\}.
\label{eqn:phiupdate}
\end{equation}
The~contribution to the update from link information,
$\nabla_{\phi_{d,n}} \mathscr{L}_{d,d'}$, depends on the choice of link
probability function. For the link probability functions
expanded in equation~(\ref{eq:approxexpect}), this term can be written as
%
\begin{equation}
\nabla_{\phi_{d,n}} \mathscr{L}_{d,d'} = (\nabla_{\overline
{\bolds{\pi}}_{d_1, d_2}}
\mathscr{L}_{d,d'}) \circ\frac{\overline{\bolds{\phi}}_{d'}}{N_d}.
\label{eqn:nudge}
\end{equation}
Intuitively, equation~(\ref{eqn:nudge}) will cause a document's latent topic
assignments to be nudged in the direction of neighboring documents'
latent topic assignments. The~magnitude of this pull depends only on
$\overline{\bolds{\pi}}_{d, d'}$, that is, some measure of how close
they are already. The~corresponding gradients for the functions in equation~(\ref{eq:approxexpect})
are
\begin{eqnarray*}
\nabla_{\overline{\bolds{\pi}}_{d, d'}}\mathscr{L}^\sigma_{d,d'}
&\approx&\bigl(1 - \sigma
({\bolds{\eta}}^{\mathrm{T}} \overline{\bolds{\pi}}_{d, d'}+ \nu)\bigr)
\bolds{\eta}, \\
\nabla_{\overline{\bolds{\pi}}_{d, d'}}\mathscr{L}^\Phi_{d,d'}
&\approx&\frac{\Phi
'({\bolds{\eta}}^{\mathrm{T}} \overline{\bolds{\pi}}_{d, d'}+ \nu
)}{\Phi({\bolds{\eta}}^{\mathrm{T}} \overline{\bolds{\pi}}_{d,
d'}+ \nu)} \bolds{\eta}, \\
\nabla_{\overline{\bolds{\pi}}_{d, d'}}\mathscr{L}^e_{d,d'} &=& \bolds{\eta}.
\end{eqnarray*}

The~gradient when $\psi_N$ is the link probability function is
%
\begin{equation}
\nabla_{\phi_{d,n}}\mathscr{L}^N_{d,d'} = \frac{2}{N_d} \bolds{\eta
} \circ\biggl(\overline{\bolds{\phi}}_{d'} - \overline{\bolds{\phi
}}_{d,-n} - \frac{\mathbf{1}}{N_d}\biggr),
\label{eq:psiNgrad}
\end{equation}
where $\overline{\bolds{\phi}}_{d,-n} = \overline{\bolds{\phi}}_d
- \frac{1}{N_d}\phi_{d,n}$.
Similar in spirit to equation~(\ref{eqn:nudge}), equation~(\ref
{eq:psiNgrad}) will cause a
document's latent topic assignments to be drawn toward those of its
neighbors. This draw is tempered by $\overline{\bolds{\phi
}}_{d,-n}$, a measure
of how similar the current document is to its neighbors.

The~contribution to the update in equation~(\ref{eqn:phiupdate}) from
the word
evidence $\log\bolds{\beta}_{\cdot, w_{d,j}}$ can be computed by taking
the element-wise logarithm of the $w_{d,j}$th column of the topic
matrix $\bolds{\beta}$. The~contribution to the update from the
document's latent topic proportions is given by
\[
\mathbb{E}_{q}[\log\bolds{\theta}_d | \bolds{\gamma}_d] = \Psi(\bolds{\gamma
}_{d}) - \Psi\Bigl(\sum\gamma_{d,
i}\Bigr),
\]
where $\Psi$ is the digamma function. (A digamma of a vector is the
vector of digammas.) The~update for $\bolds{\gamma}$ is identical to
that in variational inference for LDA [\citet{Blei2003p1135}],
\[
\gamma_d \leftarrow\alpha+ \sum_n \bolds{\phi}_{d,n}.
\]
These updates are fully derived in Appendix~\ref{app:ascent}.

\subsection{Parameter estimation}

We fit the model by finding maximum likelihood estimates for each of
the parameters: multinomial topic vectors $\bolds{\beta}_{1:K}$ and link
function parameters $\bolds{\eta}, \nu$. Once again, this is
intractable so we turn to an approximation. We employ variational
expectation-maximization, where we iterate between optimizing the ELBO
of equation~(\ref{eqn:posterior}) with respect to the variational distribution
and with respect to the model parameters. This is equivalent to the
usual expectation-maximization algorithm [\citet{Dempster1977}], except
that the computation of the posterior is replaced by variational
inference.

Optimizing with respect to the variational distribution is described
in Section~\ref{sec:inference}. Optimizing with respect to the model
parameters is equivalent to maximum likelihood estimation with
expected sufficient statistics, where the expectation is taken with
respect to the variational distribution.

The~update for the topics matrix $\bolds{\beta}$ is
%
\begin{equation}
\beta_{k,w} \propto\sum_d \sum_n \mathbh{1}(w_{d,n} = w)\phi_{d,n,k}.
\label{eq:betaupdate}
\end{equation}
This is the same as the variational EM update for
LDA [\citet{Blei2003p1135}]. In practice, we smooth our estimates of
$\beta_{k,w}$ using pseudocount smoothing [\citet{jurafsky2008sal}]
which helps to prevent overfitting by positing a Dirichlet prior on
$\bolds{\beta}_k$.

In order to fit the parameters $\bolds{\eta}, \nu$ of the logistic
function of equation~(\ref{eq:psisigma}), we employ gradient-based optimization.
Using the approximation described in equation~(\ref{eq:approxexpect}),
we compute
the gradient of the objective given in equation~(\ref{eqn:posterior}) with
respect to these parameters,
\begin{eqnarray*}
\nabla_{\bolds{\eta}} \mathscr{L} & \approx&\sum_{(d_1, d_2)}
[ y_{d_1, d_2} -
\sigma({\bolds{\eta}}^{\mathrm{T}} \overline{\bolds{\pi
}}_{d_1, d_2}+ \nu)
] \overline{\bolds{\pi}}_{d_1, d_2}, \\ 
\frac{\partial}{\partial\nu}\mathscr{L} & \approx&\sum_{(d_1,
d_2)} [
y_{d_1, d_2} -
\sigma({\bolds{\eta}}^{\mathrm{T}} \overline{\bolds{\pi
}}_{d_1, d_2}+ \nu)
].
\end{eqnarray*}

Note that these gradients cannot be used to directly optimize the
parameters of the link probability function without negative
observations (i.e., $y_{d_1, d_2} = 0$). We address this by applying
a regularization penalty. This regularization penalty along with
parameter update procedures for the other link probability functions
are given in Appendix~\ref{app:parameters}.


\subsection{Prediction}
\label{sec:prediction}
With a fitted model, our ultimate goal is to make predictions about
new data. We describe two kinds of prediction: link prediction from
words and word prediction from links.


In link prediction, we are given a new document (i.e., a document which
is not in the training set) and its words. We are asked to predict
its links to the other documents. This requires computing
\[
p(y_{d, d'} | \mathbf{w_d}, \mathbf{w_{d'}}) =
\sum_{\mathbf{z}_{d}, \mathbf{z}_{d'}} p(y_{d, d'} | \overline{\mathbf{z}}_{d},
\overline{\mathbf{z}}_{d'}) p(\mathbf{z}_{d}, \mathbf{z}_{d'} | \mathbf{w_d},
\mathbf{w_{d'}}),
\]
an expectation with respect to a posterior that we cannot compute.
Using the inference algorithm from Section~\ref{sec:inference}, we find
variational parameters which optimize the ELBO for the given evidence,
that is, the words and links for the training documents and the words in
the test document. Replacing the posterior with this approximation
$q(\bolds{\Theta}, \mathbf{Z})$, the predictive probability is approximated
with
%
\begin{equation}
p(y_{d, d'} | \mathbf{w_d}, \mathbf{w_{d'}}) \approx
\mathbb{E}_{q}[p(y_{d, d'} | \overline{\mathbf{z}}_{d}, \overline
{\mathbf{z}}_{d'})].
\label{eqn:link-prediction}
\end{equation}
In a variant of link prediction, we are given a new set of documents
(documents not in the training set) along with their words and asked
to select the links most likely to exist. The~predictive probability
for this task is proportional to equation~(\ref{eqn:link-prediction}).

The~second predictive task is word prediction, where we predict the
words of a new document based only on its links. As with link
prediction, $p(w_{d, i} | \mathbf{y_d})$ cannot be computed. Using the
same technique, a variational distribution can approximate this
posterior. This yields the predictive probability
\begin{eqnarray*}
p(w_{d, i} | \mathbf{y_d}) \approx\mathbb{E}_{q}[p(w_{d, i} | z_{d,i})].
\end{eqnarray*}

Note that models which treat the endpoints of links as discrete
observations of data indices cannot participate in the two tasks
presented here. They cannot make meaningful predictions for documents
that do not appear in the training set [\citet{Cohn2001p1231};
\citet{Erosheva2004p8486}; \citet{Nallapati2008p1331}; \citet{Sinkkonen2008p5772}]. By
modeling both documents and links generatively, our model is able to
give predictive distributions for words given links, links given
words, or any mixture thereof.

\section{Empirical results}
\label{sec:results}

We examined the RTM on four data sets.\footnote{An implementation of
the RTM with accompanying data can be found at
\url{http://cran.r-project.org/web/packages/lda/}.} Words were
stemmed; stop words, that is, words like ``and,'' ``of,'' or ``but,'' and
infrequently occurring words were removed. Directed links were
converted to undirected links\footnote{The~RTM can be extended to
accommodate directed connections. Here we modeled undirected
links.} and documents with no links were removed. The~\textit{Cora}
data [\citet{McCallum2000p5994}] contains abstracts from the Cora
computer science research paper search engine, with links between
documents that cite each other. The~\textit{WebKB}
data [\citet{Craven1998p6052}] contains web pages from the computer
science departments of different universities, with links determined
from the hyperlinks on each page. The~\textit{PNAS} data contains
recent abstracts from the Proceedings of the National Academy of
Sciences. The~links between documents are intra-\textit{PNAS}
citations. The~\textit{LocalNews} data set is a corpus of local news
culled from various media markets throughout the United States. We
create one bag-of-words document associated with each state (including
the District of Columbia); each state's ``document'' consists of
headlines and summaries from local news in that state's media markets.
Links between states were determined by geographical adjacency.
Summary statistics for these data sets are given in
Table~\ref{table:corpora}.

\begin{table}
\caption{Summary statistics for the four data sets after processing}
\label{table:corpora}
\begin{tabular*}{\textwidth}{@{\extracolsep{\fill}}lcccc@{}}
\hline
\textbf{Data set} & \textbf{\# of documents} & \textbf{\# of words} & \textbf{Number of links} & \textbf{Lexicon size} \\
\hline
\textit{Cora} & 2708 & \phantom{0,}49216 & 5278 & 1433 \\
\textit{WebKB} & \phantom{0}877 & \phantom{0,}79365 & 1388 & 1703 \\
\textit{PNAS} & 2218 & 11,9162 & 1577 & 2239 \\
\textit{LocalNews} & \phantom{00}51 & \phantom{0,}93765 & \phantom{0}107 & 1242 \\
\hline
\end{tabular*}
\end{table}


\subsection{Evaluating the predictive distribution}
\label{sec:eval-prediction}
As with any probabilistic\break model, the RTM defines a probability
distribution over unseen data. After inferring the latent variables
from data (as described in Section~\ref{sec:inference}), we ask how
well the model predicts the links and words of unseen nodes. Models
that give higher probability to the unseen documents better capture
the joint structure of words and links.

We study the RTM with three link probability functions discussed
above: the logistic link probability function, $\psi_\sigma$, of
equation~(\ref{eq:psisigma}); the exponential link probability function,
$\psi_e$, of equation~(\ref{eq:psie}); and the probit link probability function,
$\psi_\Phi$, of equation~(\ref{eq:psiphi}). We compare these models
against two
alternative approaches.\looseness=1

The~first (``Pairwise Link-LDA'') is the model proposed by
\citet{Nallapati2008p7255}, which is an extension of the mixed
membership stochastic block model [\citet{Airoldi2007p1863}] to model
network structure and node attributes. This model posits that each
link is generated as a function of two individual topics, drawn from
the topic proportions vectors associated with the endpoints of the
link. Because latent topics for words and links are drawn
independently in this model, it cannot ensure that the discovered
topics are representative of both words and links simultaneously.
Additionally, this model introduces additional variational parameters
for every link which adds computational complexity.

The~second (``LDA $+$ Regression'') first fits an LDA model to the
documents and then fits a logistic regression model to the observed
links, with input given by the Hadamard product of the latent class
distributions of each pair of documents. Rather than performing
dimensionality reduction and regression simultaneously, this method
performs unsupervised dimensionality reduction first, and then
regresses to understand the relationship between the latent space and
underlying link structure. All models were fit such that the
total mass of the Dirichlet hyperparameter $\alpha$ was 1.0. (While
we omit a full sensitivity study here, we observed that the
performance of the models was similar for $\alpha$ within a factor of
2 above and below the value we chose.)

\begin{figure}

\includegraphics{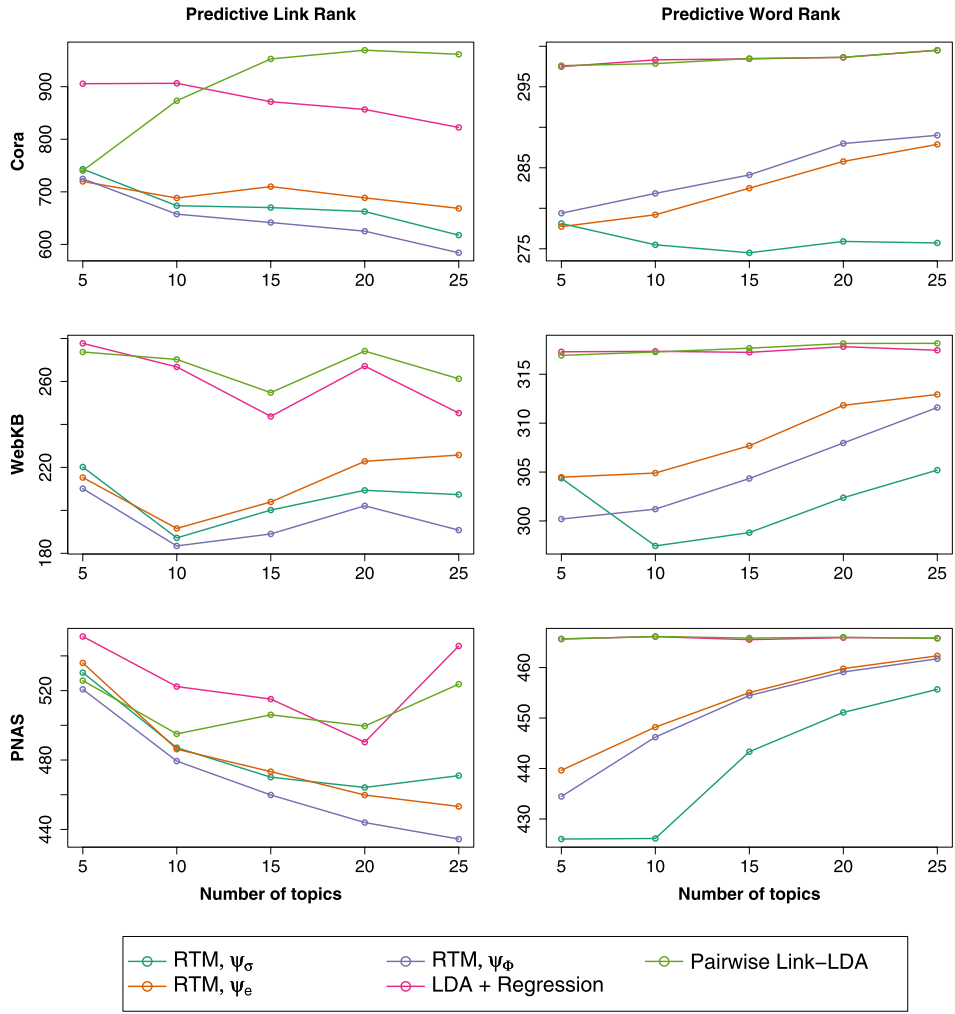}

\caption{Average held-out predictive link rank (left) and word rank
(right) as a function of the number of topics. Lower is better.
For all three corpora, RTMs outperform baseline unigram, LDA and
``Pairwise Link-LDA'' Nallapati et al. (\protect\citeyear{Nallapati2008p7255}).}
\label{figure:likelihoods}
\end{figure}

We measured the performance of these models on link prediction and
word prediction (see Section~\ref{sec:prediction}). We divided the
\textit{Cora}, \textit{WebKB} and \textit{PNAS} data sets each into five
folds. For each fold and for each model, we ask two predictive
queries: given the words of a new document, how probable are its
links; and given the links of a new document, how probable are its
words? Again, the predictive queries are for completely new test
documents that are not observed in training. During training the test
documents are removed along with their attendant links. We show the
results for both tasks in terms of predictive rank as a function of
the number of topics in Figure~\ref{figure:likelihoods}. (See
Section~\ref{sec:discussion} for a discussion on potential approaches for
selecting the number of topics and the Dirichlet hyperparameter
$\alpha$.) Here we follow the convention that lower predictive rank
is better.

In predicting links, the three variants of the RTM perform better than
all of the alternative models for all of the data sets (see
Figure~\ref{figure:likelihoods}, left column). \textit{Cora} is paradigmatic,
showing a nearly 40\% improvement in predictive rank over baseline and
25\% improvement over LDA $+$ Regression. The~performance for the RTM
on this task is similar for all three link probability functions. We
emphasize that the links are predicted to documents seen in the
training set from documents which were held out. By incorporating
link and node information in a joint fashion, the model is able to
generalize to new documents for which no link information was
previously known.

Note that the performance of the RTM on link prediction generally
increases as the number of topics is increased (there is a slight
decrease on WebKB). In contrast, the performance of the Pairwise
Link-LDA worsens as the number of topics is increased. This is most
evident on \textit{Cora}, where Pairwise Link-LDA is competitive with
RTM at five topics, but the predictive link rank monotonically
increases after that despite its increased dimensionality (and
commensurate increase in computational difficulty). We hypothesize
that Pairwise Link-LDA exhibits this behavior because it uses some
topics to explain the words observed in the training set, and other
topics to explain the links observed in the training set. This
problem is exacerbated as the number of topics is increased, making it
less effective at predicting links from word observations.

In predicting words, the three variants of the RTM again outperform
all of the alternative models (see Figure~\ref{figure:likelihoods}, right
column). This is because the RTM uses link information to influence
the predictive distribution of words. In contrast, the predictions of
LDA $+$ Regression and Pairwise Link-LDA barely use link information;
thus, they give predictions independent of the number of topics similar
to those made by a simple unigram model.



\subsection{Automatic link suggestion}

A natural real-world application of link prediction is to suggest
links to a user based on the text of a document. One might suggest
citations for an abstract or friends for a user in a social network.

As a complement to the quantitative evaluation of link prediction
given in the previous section, Table~\ref{figure:cora-qual2} illustrates
suggested citations using RTM ($\psi_e$) and LDA $+$ Regression as
\begin{table}
\caption{Top eight link predictions made by RTM ($\psi_e$) and LDA $+$
Regression for
two documents (italicized) from \textit{Cora}.
The~models were fit with 10 topics. Boldfaced titles indicate actual
documents cited by or citing each document. Over the whole
corpus, RTM improves precision over LDA $+$ Regression by
80\% when evaluated on the first 20 documents retrieved}
\label{figure:cora-qual2}
\begin{tabular*}{\textwidth}{@{\extracolsep{\fill}}cl@{}}
\hline
&\textit{Markov chain Monte Carlo convergence diagnostics: A comparative review}  \\
\hline
\multirow{8}{*}{
\begin{turn}{90}\hspace{0pt}\textbf{RTM ($\psi_e$)}
\end{turn}}
&\textbf{Minorization conditions and convergence rates for Markov chain Monte Carlo}   \\
 &Rates of convergence of the Hastings and Metropolis algorithms \\
&\textbf{Possible biases induced by MCMC convergence diagnostics}  \\
&Bounding convergence time of the Gibbs sampler in Bayesian image restoration  \\
&Self regenerative Markov chain Monte Carlo  \\
&Auxiliary variable methods for Markov chain Monte Carlo with applications  \\
 &\textbf{Rate of Convergence of the Gibbs Sampler by Gaussian Approximation} \\
&Diagnosing convergence of Markov chain Monte Carlo algorithms  \\
[5pt]
\multirow{9}{*}{
\begin{turn}{90}\hspace{-0pt}\textbf{LDA $+$ Regression}
\end{turn}}&Exact Bound for the Convergence of Metropolis Chains   \\
 &Self regenerative Markov chain Monte Carlo \\
&\textbf{Minorization conditions and convergence rates for Markov chain Monte Carlo}  \\
&Gibbs--Markov models  \\
&Auxiliary variable methods for Markov chain Monte Carlo with applications  \\
&Markov Chain Monte Carlo Model Determination for Hierarchical and Graphical Models  \\
&Mediating instrumental variables  \\
&A qualitative framework for probabilistic inference  \\
&Adaptation for Self Regenerative MCMC  \\
\hline
\end{tabular*}

\begin{tabular*}{\textwidth}{@{\extracolsep{\fill}}cl@{}}
\hline
&\textit{Competitive environments evolve better solutions for complex tasks}  \\
\hline
\multirow{8}{*}{
\begin{turn}{90}\hspace{0pt}\textbf{RTM ($\psi_e$)}
\end{turn}}&\textbf{Coevolving High Level Representations} \\
&A Survey of Evolutionary Strategies  \\
&\textbf{Genetic Algorithms in Search, Optimization and Machine Learning} \\
&\textbf{Strongly typed genetic programming in evolving cooperation strategies}  \\
&Solving combinatorial problems using evolutionary algorithms  \\
&A promising genetic algorithm approach to job-shop scheduling\ldots \\
&Evolutionary Module Acquisition  \\
&An Empirical Investigation of Multi-Parent Recombination Operators\ldots \\
[5pt]
\multirow{8}{*}{
\begin{turn}{90}\hspace{0pt}\textbf{LDA $+$ Regression}
\end{turn}}&A New Algorithm for DNA Sequence Assembly   \\
&Identification of protein coding regions in genomic DNA  \\
&Solving combinatorial problems using evolutionary algorithms  \\
&A promising genetic algorithm approach to job-shop scheduling\ldots \\
&A genetic algorithm for passive management  \\
&The~Performance of a Genetic Algorithm on a Chaotic Objective Function \\
&Adaptive global optimization with local search  \\
&Mutation rates as adaptations  \\
\hline
\end{tabular*}
\end{table}
predictive models. These suggestions were computed from a model fit
on one of the folds of the \textit{Cora} data using 10 topics.
(Results are qualitatively similar for models fit using different
numbers of topics; see Section~\ref{sec:discussion} for strategies for
choosing the number of topics.) The~top results illustrate suggested
links for ``Markov chain Monte Carlo convergence diagnostics: A
comparative review,'' which occurs in this fold's training set. The~bottom results illustrate suggested links for ``Competitive
environments evolve better solutions for complex tasks,'' which is in
the test set.

RTM outperforms LDA $+$ Regression in being able to identify more true
connections. For the first document, RTM finds 3 of the connected
documents versus~1 for LDA $+$ Regression. For the second document, RTM
finds 3 while LDA $+$ Regression does not find any. This qualitative
behavior is borne out quantitatively over the entire corpus.
Considering the precision of the first 20 documents retrieved by the
models, RTM improves precision over LDA $+$ Regression by 80\%. (Twenty
is a reasonable number of documents for a user to examine.)

While both models found several connections which were not observed in
the data, those found by the RTM are qualitatively different. In the
first document, both sets of suggested links are about Markov chain
Monte Carlo. However, the RTM finds more documents relating
specifically to convergence and stationary behavior of Monte Carlo
methods. LDA $+$ Regression finds connections to documents in the
milieu of MCMC, but many are only indirectly related to the input
document. The~RTM is able to capture that the notion of
``convergence'' is an important predictor for citations, and has
adjusted the topic distribution and predictors correspondingly. For
the second document, the documents found by the RTM are also of a
different nature than those found by LDA $+$ Regression. All of the
documents suggested by RTM relate to genetic algorithms. LDA $+$
Regression, however, suggests some documents which are about genomics.
By relying only on words, LDA $+$ Regression conflates two ``genetic''
topics which are similar in vocabulary but different in citation
structure. In contrast, the RTM partitions the latent space
differently, recognizing that papers about DNA sequencing are unlikely
to cite papers about genetic algorithms, and vice versa. Better
modeling the properties of the network jointly with the content of the
documents, the model is able to better tease apart the community
structure.

\subsection{Modeling spatial data}

While explicitly linked structures like citation networks offer one
sort of connectivity, data with spatial or temporal information offer
another sort of connectivity. In this section we show how RTMs can
be used to model spatially connected data by applying it to the
\textit{LocalNews} data set, a corpus of news headlines and summaries
from each state, with document linkage determined by spatial
adjacency.

\begin{figure}

\includegraphics{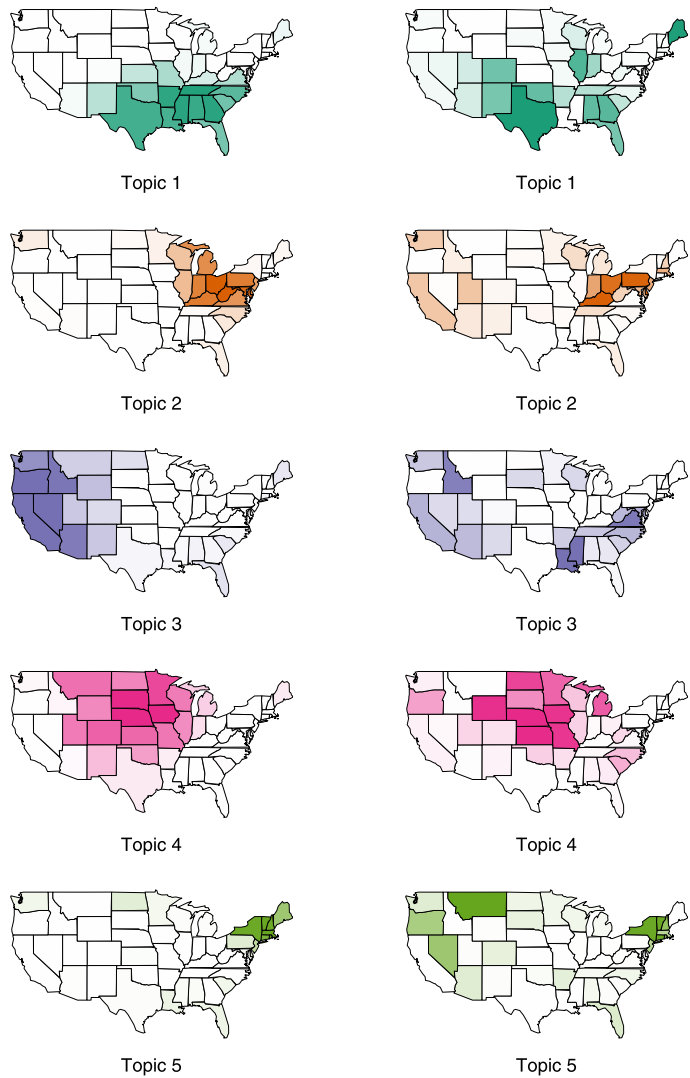}

\caption{A comparison between RTM (left) and LDA (right) of topic
distributions on local news data. Each color/row depicts a single
topic. Each state's color intensity indicates the magnitude of
that topic's component. The~corresponding words associated with
each topic are given in Table~\protect\ref{table:statewords}. Whereas LDA
finds geographically diffuse topics, RTM, by modeling spatial
connectivity, finds coherent regions.}
\label{fig:states}
\end{figure}

Figure~\ref{fig:states} shows the per state topic distributions
inferred by
RTM (left) and LDA (right). Both models were fit with five topics
using the same initialization. (We restrict the discussion here to
five topics for expositional convenience. See Section~\ref{sec:discussion}
for a discussion on potential approaches for selecting the number of
topics.) While topics are, strictly speaking, exchangeable and
therefore not comparable between models, using the same initialization
typically yields topics which are amenable to comparison. Each row of
Figure~\ref{fig:states} shows a single component of each state's topic
proportion for RTM and LDA. That is, if $\theta_s$ is the latent
topic proportions vector for state $s$, then $\theta_{s1}$ governs the
intensity of that state's color in the first row, $\theta_{s2}$~the
second, and so on.

While both RTM and LDA model the words in each state's local news
corpus, LDA ignores geographical information. Hence, it finds topics
which are distributed over a wide swath of states which are often not
contiguous. For example, LDA's topic~1 is strongly expressed by Maine
and Illinois, along with Texas and other states in the South and West.
In contrast, RTM only assigns nontrivial mass to topic~1 in
Southern states. Similarly, LDA finds that topic 5 is expressed by
several states in the Northeast and the West. The~RTM, however,
concentrates topic 4's mass on the Northeastern states.

\begin{table}
\caption{The~top eight words in each RTM (left) and LDA (right) topic
shown in Figure~\protect\ref{fig:states} ranked by score (defined below). RTM
finds words which are predictive of both a state's geography and
its~local~news}
\label{table:statewords}
\begin{tabular}{c}

\includegraphics{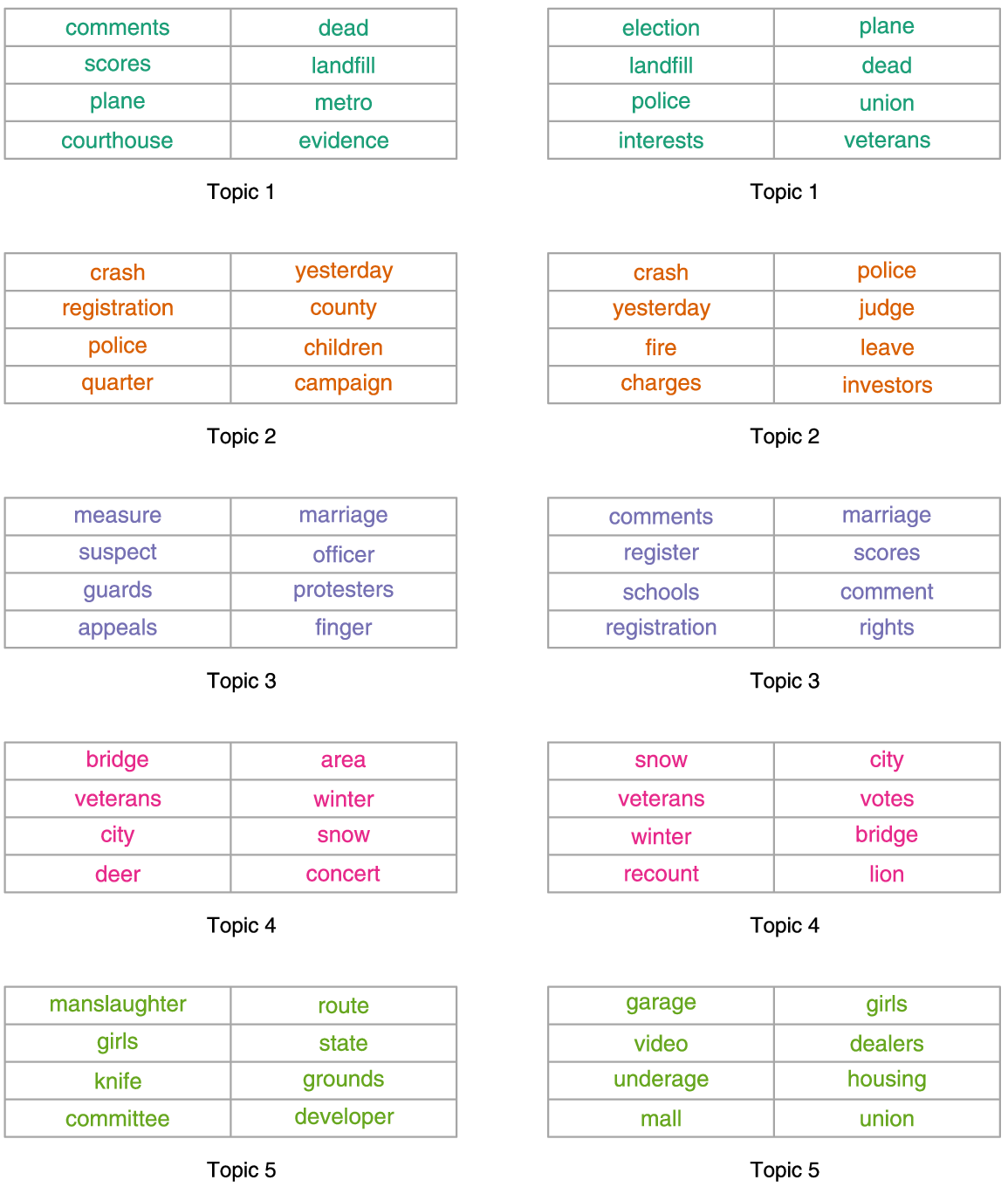}

\end{tabular}\vspace*{-6pt}
\end{table}


The~RTM does so by finding different topic assignments for each state
and, commensurately, different distributions over words for each topic.
Table~\ref{table:statewords} shows the top words in each RTM topic and
each LDA topic. Words are ranked by the following score:
\[
\mathrm{score}_{k,w} \equiv\beta_{k,w} \biggl(\log\beta_{k,w} - \frac
{1}{K} \sum_{k'} \log\beta_{k', w}\biggr).
\]
The~score finds words which are likely to appear in a topic, but also
corrects for frequent words. The~score therefore puts greater weight
on words which more easily characterize a topic.
Table~\ref{table:statewords} shows that RTM finds words more
geographically indicative. While LDA provides one way of analyzing
this collection of documents, the RTM enables a different approach
which is geographically cognizant. For example, LDA's topic 3 is an
assortment of themes associated with California (e.g., ``marriage'') as
well as others (``scores,'' ``registration,'' ``schools''). The~RTM, on the
other hand, discovers words thematically related to a single news item
(``measure,'' ``protesters,'' ``appeals'') local to California. The~RTM
typically finds groups of words associated with specific news stories,
since they are easily localized, while LDA finds words which cut
broadly across news stories in many states. Thus, on topic 5, the RTM
discovers key words associated with news stories local to the
Northeast such as ``manslaughter'' and ``developer.'' On topic 5, the RTM
also discovers a peculiarity of the Northeastern dialect: that roads
are given the appellation ``route'' more frequently than elsewhere in
the country.


By combining textual information along with geographical information,
the RTM provides a novel exploratory tool for identifying clusters of
words that are driven by both word co-occurrence and geographic
proximity. Note that the RTM finds regions in the United States which
correspond to typical clusterings of states: the South, the Northeast,
the Midwest, etc. Further, the soft clusterings found by RTM confirm
many of our cultural intuitions---while New York is definitively a
Northeastern state, Virginia occupies a liminal space between the
MidAtlantic and the South.


\section{Discussion}
\label{sec:discussion}
There are many avenues for future work on relational topic models.
Applying the RTM to diverse types of ``documents'' such as
protein-interaction networks or social networks, whose node attributes
are governed by rich internal structure, is one direction. Even the
text documents which we have focused on in this paper have internal
structure such as syntax [\citet{BoydGraber2008}] which we are
discarding in the bag-of-words model. Augmenting and specializing the
RTM to these cases may yield better models for many application
domains.

As with any parametric mixed-membership model, the number of latent
components in the RTM must be chosen using either prior knowledge or
model-selection techniques such as cross-validation. Incorporating
non-parametric Bayesian priors such as the Dirichlet process into the
model would allow it to flexibly adapt the number of topics to the
data [\citet{Ferguson1973}; \citet{Antoniak1974}; \citet{Kemp2004p2068}; \citet{Teh2007}].
This, in turn, may give researchers new insights into the latent
membership structure of networks.

In sum, the RTM is a hierarchical model of networks and per-node
attribute data. The~RTM is used to analyze linked corpora such as
citation networks, linked web pages, social networks with user
profiles, and geographically tagged news. We have demonstrated
qualitatively and quantitatively that the RTM provides an effective
and useful mechanism for analyzing and using such data. It
significantly improves on previous models, integrating both
node-specific information and link structure to give better
predictions.

\begin{appendix}

\section{Derivation of coordinate ascent updates}
\label{app:ascent}

Inference under the variational method amounts to finding values of
the variational parameters $\bolds{\gamma}, \bolds{\Phi}$ which optimize the
evidence lower bound, $\mathscr{L}$, given in equation~(\ref{eqn:posterior}).
To do so, we first expand the expectations in these terms:
\begin{eqnarray}\label{eq:elbo}
\mathscr{L} &=& \sum_{(d_1, d_2)} \mathscr{L}_{d_1, d_2} + \sum_d
\sum_n {\phi_{d,n}}^{\mathrm{T}} \log\beta_{\cdot, w_{d,n}}
\nonumber
\\
&&{} +\sum_d \sum_n {\phi_{d,n}}^{\mathrm{T}} \bigl(\Psi(\gamma_d) -
\mathbf{1} \Psi({\mathbf{1}}^{\mathrm{T}} \gamma_d)\bigr)  \nonumber\\
& &{}+\sum_d {(\alpha- 1)}^{\mathrm{T}} \bigl(\Psi(\gamma_d) - \mathbf{1}
\Psi({\mathbf{1}}^{\mathrm{T}} \gamma_d)\bigr)  \nonumber\\[-8pt]\\[-8pt]
& &{}+\sum_d \sum_n {\phi_{d,n}}^{\mathrm{T}} \log\phi_{d,n}
\nonumber\\
& &{}-\sum_d {(\gamma_d - 1)}^{\mathrm{T}} \bigl(\Psi(\gamma_d) - \mathbf
{1} \Psi({\mathbf{1}}^{\mathrm{T}} \gamma_d)\bigr)  \nonumber\\
& &{}+\sum_d {\mathbf{1}}^{\mathrm{T}} \log\Gamma(\gamma_d) - \log
\Gamma({\mathbf{1}}^{\mathrm{T}} \gamma_d),\nonumber
\end{eqnarray}
where $\mathscr{L}_{d_1,d_2}$ is defined as in equation~(\ref{eqn:firstterm}).
Since $\mathscr{L}_{d_1, d_2}$ is independent of $\bolds{\gamma}$, we can
collect all of the terms associated with $\gamma_d$ into
\begin{eqnarray*}
\mathscr{L}_{\gamma_d} &=& {\biggl(\alpha+ \sum_n \phi _{d,n}
-\gamma_d\biggr)}^{\mathrm{T}} \bigl(\Psi(\gamma_d) - \mathbf{1} \Psi
({\mathbf{1}}^{\mathrm{T}} \gamma_d)\bigr)  \\
& &{}+{\mathbf{1}}^{\mathrm{T}} \log\Gamma(\gamma_d) - \log\Gamma
({\mathbf{1}}^{\mathrm{T}} \gamma_d).
\end{eqnarray*}
Taking the derivatives and setting equal to zero leads to the following
optimality condition:
\[
{\biggl(\alpha+ \sum_n \phi_{d,n} -\gamma_d\biggr)}^{\mathrm{T}}
\bigl(\Psi'(\gamma_d) - \mathbf{1} \Psi'({\mathbf{1}}^{\mathrm{T}} \gamma_d)\bigr)
= 0,
\]
which is satisfied by the update
%
\begin{equation}
\gamma_d \leftarrow\alpha+ \sum_n \phi_{d,n}.
\end{equation}
In order to derive the update for $\phi_{d,n}$, we also collect its
associated terms,
\[
\mathscr{L}_{\phi_{d,n}} = {\phi_{d,n}}^{\mathrm{T}} \bigl(\log\phi
_{d,n} + \log\beta_{\cdot, w_{d,n}} + \Psi(\gamma_d) - \mathbf{1}
\Psi({\mathbf{1}}^{\mathrm{T}} \gamma_d)\bigr) + \sum_{d' \ne d} \mathscr
{L}_{d, d'}.
\]
Adding a Lagrange multiplier to ensure that $\phi_{d,n}$ normalizes
and setting the derivative equal to zero leads to the following condition:
%
\begin{equation}
\phi_{d,n} \propto\exp\{\log\beta_{\cdot, w_{d,n}} + \Psi
(\gamma_d) - \mathbf{1} \Psi({\mathbf{1}}^{\mathrm{T}} \gamma_d) + \nabla
_{\phi_{d,n}} \mathscr{L}_{d,d'} \}.
\end{equation}
The~exact form of $\nabla_{\phi_{d,n}} \mathscr{L}_{d,d'}$ will depend
on the link probability function chosen. If the expected log link
probability depends only on $\overline{\bolds{\pi}}_{d_1, d_2}=
\overline{\bolds{\phi}}_{d_1} \circ
\overline{\bolds{\phi}}_{d_2}$, the gradients are given by
equation~(\ref{eqn:nudge}). When
$\psi_N$ is chosen as the link probability function, we expand the
expectation,
\begin{eqnarray}\label{eq:expandpsiN}
\hspace*{10pt}\mathbb{E}_{q}[\log\psi_N(\overline{\mathbf{z}}_{d}, \overline
{\mathbf{z}}_{d'})] &=&
-{\bolds{\eta}}^{\mathrm{T}}\mathbb{E}_{q}[(\overline{\mathbf{z}}_d
- \overline{\mathbf{z}}_{d'}) \circ(\overline{\mathbf{z}}_d -
\overline{\mathbf{z}}_{d'})] - \nu\hspace*{-10pt}\nonumber\\[-8pt]\\[-8pt]
\hspace*{10pt}&=&-\nu-\sum_i \eta_i (\mathbb{E}_{q}[\overline{\mathbf
{z}}_{d,i}^2] + \mathbb{E}_{q}[\overline{\mathbf{z}}_{d',i}^2 ]- 2
\overline{\bolds{\phi}}_{d,i}\overline{\bolds{\phi}}_{d',i}).\hspace*{-10pt}\nonumber
\end{eqnarray}
Because each word is independent under the variational distribution,\vspace*{-3pt}
$\mathbb{E}_{q}[\overline{\mathbf{z}}_{d,i}^2] = \operatorname
{Var}(\overline{\mathbf{z}}_{d,i}) +
\overline{\bolds{\phi}}_{d,i}^2$, where $\operatorname
{Var}(\overline{\mathbf{z}}_{d,i}) = \frac{1}{N_d^2}
\sum_n \phi_{d,n,i} (1 - \phi_{d,n,i})$. The~gradient of this
expression is given by equation~(\ref{eq:psiNgrad}).



\section{Derivation of parameter estimates}
\label{app:parameters}

In order to estimate the parameters of our model, we find values of
the topic multinomial parameters $\bolds{\beta}$ and link probability
parameters $\bolds{\eta}, \nu$ which maximize the variational objective,
$\mathscr{L}$, given in equation~(\ref{eqn:posterior}).

To optimize $\bolds{\beta}$, it suffices to take the derivative of the
expanded objective given in equation~(\ref{eq:elbo}) along with a Lagrange
multiplier to enforce normalization:
\[
\partial_{\beta_{k,w}} \mathscr{L} = \sum_d \sum_n \phi_{d,n,k}
\mathbh{1}(w = w_{d,n}) \frac{1}{\beta_{k,w_{d,n}}} + \lambda_k.
\]
Setting this quantity equal to zero and solving yields the update
given in equation~(\ref{eq:betaupdate}).

By taking the gradient of equation~(\ref{eq:elbo}) with respect to $\bolds{\eta}$
and $\nu$, we can also derive updates for the link probability
parameters. When the expectation of the logarithm of the link
probability function depends only on ${\bolds{\eta}}^{\mathrm{T}}
\overline{\bolds{\pi}}_{d, d'}
+ \nu$, as with all the link functions given in
equation~(\ref{eq:approxexpect}), then these derivatives take a
convenient form.
For notational expedience, denote $\bolds{\eta}^+ = \langle\bolds{\eta},
\nu\rangle$ and $\overline{\bolds{\pi}}_{d, d'}^+ = \langle
\overline{\bolds{\pi}}_{d, d'}, 1 \rangle$. Then
the derivatives can be written as
%
\begin{eqnarray}\label{eq:nablaeta}
\nabla_{\eta^+}\mathscr{L}^\sigma_{d,d'} &\approx&\bigl(1 - \sigma
({\bolds{\eta}^{+\mathrm{T}}} \overline{\bolds{\pi}}_{d, d'}^+)\bigr)
\overline{\bolds{\pi}}_{d, d'}^+, \nonumber\\
\nabla_{\eta^+}\mathscr{L}^\Phi_{d,d'} &\approx&\frac{\Phi
'({\bolds{\eta}^{+\mathrm{T}}} \overline{\bolds{\pi}}_{d,
d'}^+)}{\Phi({\bolds{\eta }^{+\mathrm{T}}} \overline{\bolds{\pi
}}_{d, d'}^+)} \overline{\bolds{\pi}}_{d, d'}^+, \\
\nabla_{\eta^+}\mathscr{L}^e_{d,d'} &=& \overline{\bolds{\pi}}_{d, d'}^+.\nonumber
\end{eqnarray}
Note that all of these gradients are positive because we are faced
with a one-class estimation problem. Unchecked, the parameter
estimates will diverge. While a variety of techniques exist to
address this problem, one set of strategies is to add regularization.

A common regularization for regression problems is the $\ell_2$
regularizer. This penalizes the objective $\mathscr{L}$ with the term
$\lambda\|\bolds{\eta}\|_2$, where $\lambda$ is a free parameter. This
penalization has a Bayesian interpretation as a Gaussian prior on
$\bolds{\eta}$.

In lieu of or in conjunction with $\ell_2$ regularization, one can
also employ regularization which in effect injects some number of
observations, $\rho$, for which the link variable $y = 0$. We
associate with these observations a document similarity of $\overline
{\bolds{\pi}}_{\alpha}
= \frac{\bolds{\alpha}}{{\mathbf{1}}^{\mathrm{T}}\bolds{\alpha}} \circ
\frac{\bolds{\alpha}}{{\mathbf{1}}^{\mathrm{T}}\bolds{\alpha}}$, the expected
Hadamard product of any two documents given\vspace*{1pt} the Dirichlet prior of the
model. Because both $\psi_\sigma$ and $\psi_\Phi$ are symmetric,
these gradients of these regularization terms can be written as
\begin{eqnarray*}
\nabla_{\eta^+}\mathscr{R}^\sigma&=& -\rho\sigma({\bolds{\eta
}^{+\mathrm{T}}} \overline{\bolds{\pi}}_{\alpha}^+) \overline
{\bolds{\pi}}_{\alpha}^+, \\
\nabla_{\eta^+}\mathscr{R}^\Phi&=& -\rho\frac{\Phi'(-{\bolds{\eta
}^{+\mathrm{T}}} \overline{\bolds{\pi}}_{\alpha}^+)}{\Phi(-{\bolds{\eta}^{+\mathrm{T}}}
\overline{\bolds{\pi}}_{\alpha}^+)} \overline{\bolds{\pi
}}_{\alpha}^+.
\end{eqnarray*}
While this approach could also be applied to $\psi_e$, here we use a
different approximation. We do this for two reasons. First, we
cannot optimize the parameters of $\psi_e$ in an unconstrained fashion
since this may lead to link functions which are not probabilities.
Second, the approximation we propose will lead to explicit updates.

Because $\mathbb{E}_{q}[\log\psi_e(\overline{\mathbf{z}}_d \circ
\overline{\mathbf{z}}_{d'})]$ is
linear in $\overline{\bolds{\pi}}_{d, d'}$ by equation~(\ref
{eq:approxexpect}), this suggests a
linear approximation of $\mathbb{E}_{q}[\log(1 - \psi_e(\overline
{\mathbf{z}}_d \circ \overline{\mathbf{z}}_{d'}))]$. Namely, we let
\[
\mathbb{E}_{q}\bigl[\log\bigl(1 - \psi_e(\overline{\mathbf{z}}_d \circ
\overline{\mathbf{z}}_{d'})\bigr)\bigr] \approx{\bolds{\eta}'}^{\mathrm
{T}}\overline{\bolds{\pi}}_{d, d'}+ \nu'.
\]
This leads to a penalty term of the form
\[
\mathscr{R}^e = \rho( {\bolds{\eta}'}^{\mathrm{T}}\overline{\bolds
{\pi}}_{\alpha}+ \nu').
\]
We fit the parameters of the approximation, $\bolds{\eta}', \nu'$, by
making the approximation exact whenever $\overline{\bolds{\pi}}_{d,
d'}= \mathbf{0}$ or $\max
\overline{\bolds{\pi}}_{d, d'}= 1$. This yields the following $K+1$
equations for the
$K+1$ parameters of the approximation:
\begin{eqnarray*}
\nu' &=& \log\bigl(1 - \exp(\nu)\bigr), \\
\eta'_i &=& \log\bigl(1 - \exp(\eta_i + \nu)\bigr) - \nu'.
\end{eqnarray*}
Combining the gradient of the likelihood of the observations given in
equation~(\ref{eq:nablaeta}) with the gradient of the penalty $\mathscr{R}^e$
and solving leads to the following updates:
\begin{eqnarray*}
\nu&\gets&\log(M - {\mathbf{1}}^{\mathrm{T}} \bar{\bolds{\Pi
}})
- \log\bigl(\rho(1 - {\mathbf{1}}^{\mathrm{T}}\overline{\bolds{\pi
}}_{\alpha}) + M -
{\mathbf{1}}^{\mathrm{T}} \bar{\bolds{\Pi}}\bigr), \\
\eta&\gets&\log( \bar{\bolds{\Pi}}) - \log(
\bar{\bolds{\Pi}} + \rho\overline{\bolds{\pi}}_{\alpha}) -
\mathbf{1}\nu,
\end{eqnarray*}
where $M = \sum_{(d_1, d_2)} 1$ and $\bar{\bolds{\Pi}} = \sum_{(d_1,
d_2)} \overline{\bolds{\pi}}_{d_1, d_2}$. Note that because of the
constraints on our
approximation, these updates are guaranteed to yield parameters for
which $0 \le\psi_e \le1$.

Finally, in order to fit parameters for $\psi_N$, we begin by
assuming the variance terms of equation~(\ref{eq:expandpsiN}) are small.
equation~(\ref{eq:expandpsiN}) can then be written as
\[
\mathbb{E}_{q}[\log\psi_N(\overline{\mathbf{z}}_{d}, \overline
{\mathbf{z}}_{d'})] =
-\nu-{\bolds{\eta}}^{\mathrm{T}} (\overline{\bolds{\phi}}_{d} -
\overline{\bolds{\phi}}
_{d'})\circ(\overline{\bolds{\phi}}_{d} - \overline{\bolds{\phi}}_{d'}),
\]
which is the log likelihood of a Gaussian distribution where
$\overline{\bolds{\phi}}_{d} - \overline{\bolds{\phi}}_{d'}$ is
random with mean 0 and
diagonal variance $\frac{1}{2\bolds{\eta}}$. This suggests fitting
$\bolds{\eta}$ using the empirically observed variance:
\[
\bolds{\eta} \gets\frac{M}{2 \sum_{d,d'} (\overline{\bolds{\phi
}}_{d} -
\overline{\bolds{\phi}}_{d'}) \circ(\overline{\bolds{\phi}}_{d}
- \overline{\bolds{\phi}}_{d'})}.
\]
$\nu$ acts as a scaling factor for the Gaussian distribution; here we
want only to ensure that the total probability mass respects the
frequency of observed links to regularization ``observations.''
Equating the normalization constant of the distribution with the
desired probability mass yields the update
\[
\nu\gets\log\tfrac{1}{2}\pi^{K/2} + \log(\rho+ M) - \log M -
\tfrac{1}{2}{\mathbf{1}}^{\mathrm{T}}\log\eta,
\]
guarding against values of $\nu$ which would make $\psi_N$
inadmissable as a probability.
\end{appendix}

\printaddresses
\end{document}